\documentclass{article}
\usepackage{ams}
\title{GRAVITATIONAL POTENTIAL ENERGY GROUP THEORETICALLY}
\author{Joachim Nzotungicimpaye\footnote{On absence from the University of Burundi}\\Kigali Institute of Education, Department of Mathematics\\ P.O.Box 5039,Kigali-Rwanda\\e-mail kimpaye @kie.ac.rw}

\begin{document}
\maketitle
\begin{abstract}
We show by symplectically realizing the one spatial Aristotle Lie
group that the hamiltonian of the associated elementary system
consist of a gravitational potential energy only.No kinetic term.
\end{abstract}

 \section{One spatial Aristotle Lie group $A$}
 We know that the Aristotle Lie group \cite{sou} $A(n),1\leq n \leq 3 $ also called static group \cite{lev},is the
space-time transformations group
\begin{equation}
x^{\prime i}_0=R^i_jx^j_0+x^i; t^\prime_0=t_0+t
\end{equation}
 The multiplication law for this group is
\begin{equation}
(t,x^i,R^i_k)(t^\prime,x^{\prime k},R^{\prime
k}_j)=(t+t^\prime,R^i_k x^{\prime k}+x^i,R^i_k R^{\prime k}_j)
\end{equation}
 One sees that the Aristotle Lie group $A(n)$ is a direct product of
 the euclidean Lie group $E(n)$ and the time translation group. In
 this paper we interest us on the one spatial dimensional Lie
 group that we will simply denote by $A$.It is the one dimensional space-time translation Lie
 group.Its multiplication law is
\begin{equation}
(t,h)(t^\prime,h^\prime)=(t+t^\prime,h+h^\prime)
\end{equation}
It is an abelian additive group . The central extension $\cal{G}$
of its Lie algebra is generated by $P, E$ and $M$ such that the
non trivial Lie brackets are \cite{ham}
\begin{equation}\label{eq:algext}
[P,E]=gM
\end{equation}
$g$ being a constant whose the physical interpretation will be
given below.If the general element of the connected Lie group $G$
generated by $P, E$ and $M$ is written as $exp(\xi M)exp(xP+tE)$ ,
then the multiplication law for $G$ is \cite{ham}
\begin{equation}\label{eq:ext}
(\xi,t,h)(\xi^\prime,t^\prime,h^\prime)=(\xi + \xi ^\prime +
ght^\prime, t+ t^\prime, h + h^\prime)
\end{equation}
\section{Coadjoint orbit of $A$}
If we denote by $\cal{G^*}$ the dual of $\cal{G}$ and if the
duality is defined by
\begin{equation}\label{eq:dua}
<(m,e,p),(\delta \xi,\delta t, \delta x>=m \delta \xi + e \delta t
+ p \delta x
\end{equation}
we find that the coadjoint action of the Aristotle Lie on
$\cal{G^*}$ is
\begin{equation}
Ad^*_{(t,h)}(m,e,p)=(m,e-mgh,p+mgt)
\end{equation}
The coadjoint orbit is then characterized by the invaraint m.Let
us denote it by ${\cal O} $ .From (\ref{eq:algext}) one verify
that the symplectic form on ${\cal O} $ is
\begin{equation}
\sigma =dp \wedge dq
\end{equation}
where $q=- \frac{e}{mg}$ and that the canonical action of the
Aristotle Lie group is then
\begin{equation}\label{eq:canon}
\phi_{(t,h)}(p, q)=(p+mgt,q + h)
\end{equation}
The Aristotelian Lie algebra is then represented on ${\cal O} $ by
the hamiltonian vector fields
\begin{equation}
\phi_*(P)=-\frac{\partial}{\partial q}; \phi _*(E)=-mg
\frac{\partial}{\partial p}
\end{equation}
One then can verify that the comomentum \cite{carinena} $\lambda
:\cal {A}\longrightarrow C^\infty ({\cal} O ,\Re)$ is such that
\begin{equation}
\lambda (E)=-mgq ; \lambda (P)=p
\end{equation}
If we interpret $g$ as the gravitational acceleration, then by
(\ref{eq:ext}) one can verify that the physical dimension for
$\xi$ is $L^2 T^{-1}$ where $L$ and $T$ are respectively the
symbols for length and time. Moreover, if the physical dimension
of (\ref{eq:dua}) is that of an action $(M L^2 T^{-1})$ , then the
physical dimensions for $m,e$ and $p$ are respectively mass ,
energy and linear momentum . From (\ref{eq:canon}) one sees that
the evolution law is given by
\begin{equation}\label{eq:evolu}
\phi_{(t,0)}(p(0), q(0))=(p(0)+mgt,q(0))
\end{equation}
and that the generator of time translations is then represented by
the dynamical vector field
\begin{equation}\label{eq:vectham}
X_H=\frac{\partial}{\partial t}- mg \frac{\partial}{\partial p}
\end{equation}
 from which one can verify that the hamiltonian is
\begin{equation}
H=mgq
\end{equation}
We recognize in it the gravitational potential energy .From
(\ref{eq:vectham}) one sees that the orbit is a static elementary
particle of mass $m$ ,linear momentum $p$ in the potential $V=mgq$
\cite{any}.

\end{document}